\documentclass{article}

\PassOptionsToPackage{numbers, compress}{natbib}

  \usepackage[final]{neurips_2019}

\usepackage{wrapfig,lipsum,booktabs}

\usepackage[utf8]{inputenc} 
\usepackage[T1]{fontenc}    
\usepackage{hyperref}       
\usepackage{url}            
\usepackage{booktabs}       
\usepackage{amsfonts}       
\usepackage{nicefrac}       
\usepackage{microtype}      
\usepackage{subfig}
\usepackage{amsmath} 
\usepackage{amssymb} 
\usepackage{array}
\usepackage{float}
\usepackage{enumitem}
\usepackage{graphicx}
\usepackage{natbib} 
\usepackage{booktabs} 
\usepackage{easytable}
\usepackage{syntonly}
\usepackage{layout}
\usepackage{pstricks}
\usepackage{tikz}
\usepackage{natbib}
\usepackage{color,soul}

\newcommand{\aacomment}[1]{\textcolor{blue}{{\bf Anima:}  #1}}
\newcommand{\tancomment}[1]{\textcolor{magenta}{{\bf Tan:}  #1}}

\newcommand*\samethanks[1][\value{footnote}]{\footnotemark[#1]}

\title{Turbulence forecasting via Neural ODE}

\author{%
 Gavin D. Portwood\thanks{Authors contributed equally}\\
 Los Alamos National Laboratory\\
  Los Alamos, NM 87545\\
  \texttt{portwood@lanl.gov} \\
  \And
    Peetak P.~Mitra\samethanks[1] \hspace{0.1pt} \thanks{Corresponding author: pmitra@umass.edu} \\
  Los Alamos National Laboratory\\
  Los Alamos, NM 87545 \\
  \texttt{peetak@lanl.gov} \\
  \And
Mateus Dias Ribeiro\samethanks[1]\\
DFKI GmbH\\
 Kaiserslautern, Germany \\
 \texttt{mateus.dias\_ribeiro@dfki.de} \\
    \And
   Tan Minh Nguyen \\
NVIDIA Corporation \\
  Santa Clara, CA 95051\\
  \texttt{mn15@rice.edu} \\
    \And
   Balasubramanya T. Nadiga \\
Los Alamos National Laboratory\\
  Los Alamos, NM 87545 \\
  \texttt{balu@lanl.gov} \\
    \And
   Juan A. Saenz \\
 Los Alamos National Laboratory\\
  Los Alamos, NM 87545 \\
  \texttt{juan.saenz@lanl.gov} \\
    \And
   Michael Chertkov \\
 Program in Applied Mathematics,\\
  University of Arizona, AZ 85719 \\
  \texttt{chertkov@math.arizona.edu} \\
    \And
   Animesh Garg \\
   University of Toronto,\\
  Toronto, ON M5S, Canada \\
  \texttt{garg@cs.toronto.edu} \\    
   \And
    Anima Anandkumar \\
 NVIDIA Corporation\\
  Santa Clara, CA 95051\\
  \texttt{anima@caltech.edu} \\
  \And
    Andreas Dengel \\
DFKI GmbH\\
  Kaiserslautern, Germany\\
  \texttt{andreas.dengel@dfki.de} \\
  \And
   Richard Baraniuk \\
  Rice University\\
  Houston, TX 77005 \\
  \texttt{richb@rice.edu} \\
    \And
   David P. Schmidt \\
  University of Massachusetts\\
  Amherst, MA 01003 \\
  \texttt{schmidt@acad.umass.edu} \\
}

\begin{document}

\maketitle

\begin {abstract}

Fluid turbulence is characterized by strong coupling across a broad range of scales. Furthermore, besides the usual local cascades, such coupling may extend to interactions that are non-local in scale-space. As such the computational demands associated with explicitly resolving the full set of scales and their interactions, as in the Direct Numerical Simulation (DNS) of the Navier-Stokes equations, in most problems of practical interest are so high that reduced modeling of scales and interactions is required before further progress can be made. While popular reduced models are typically based on phenomenological modeling of relevant turbulent processes, recent advances in machine learning techniques have energized efforts to further improve the accuracy of such reduced models. In contrast to such efforts that seek to improve an existing turbulence model, we propose a machine learning (ML) methodology that captures, de novo, underlying turbulence phenomenology without a pre-specified model form. To illustrate the approach, we consider transient modeling of the dissipation of turbulent kinetic energy---a fundamental turbulent process that is central to a wide range of turbulence models---using a Neural ODE approach. After presenting details of the methodology, we show that this approach out-performs state-of-the-art approaches. 
\end{abstract}

\section{Introduction}
Forecasting the evolution of turbulence is a critical necessity for applications from engineering design to climate modeling.
The range of scales involved in a turbulent flow is characterized by the Reynolds number, $\mathrm{Re}$. Realistic turbulent flows often have moderate to large $\mathrm{Re}$ and involve a wide range of scales, i.e. large scale separation. Scale separation arises from non-linear interactions between flow components and processes that are often non-local in space and time. The dynamics at different scales are often closely coupled and for a truly predictive numerical framework, need to be well-resolved. Due to the non-linearity of the physics, simulating the entire range of scales is almost always prohibitively expensive in real-world applications, and therein lies the need to formulate reduced order formulations that can accurately predict these multi-scale physics.

Direct numerical simulations (DNS) of the entire range of dynamical scale interactions is possible in a number of idealized canonical flows relevant to applications in weather, and climate modeling. However, in many engineering applications, only the large scale dynamics have practical relevance. A common modeling approach, in such a paradigm, is to then resolve (i.e. directly simulate) dynamics at large scales while modeling the non-linear dynamical interactions between the resolved and the unresolved scale. These models are commonly phenomenological and heuristic in nature, and fitting, or calibration, and validation of these models to DNS data, for such idealized flows, constitutes a major practical hurdle in many fields. This is because it is performed in a cumbersome manner that involves manually iterating the adjustment of parameters until good matches with the DNS data are observed. Given the need to tune these engineering models based on high-fidelity turbulence data, in this manuscript, we develop an automatic methodology for training a family of phenomenology-informed and properly parameterized reduced models. 

Recently, Deep Neural Network (DNN)-based approaches related to modeling fluid problems has gained wide attention~\cite{portwood2019physics, portwood2019autonomous, mitra2019data}. Prominent among these are approaches based on modeling these dynamical systems as differential equations \cite{chen2018neural,tan2018infocnf,rubanova2019latent}. These methods require replacing the residual networks (ResNet/RNN) with ordinary differential equations (ODE) \cite{chen2018neural,sun2019neupde}, where Neural ODE (NODE) has emerged as a popular approach. NODE is a supervised machine learning approach that is based on learning the latent space representations of dynamical equations. It is devised in a non-intrusive manner without a pre-specified model form, that makes this framework very appealing for complex, transient, non-linear physics problems as considered here~\cite{maulik2019time, san2019artificial, dias2019data}. 

In this work, we look at a simple system of couple ODEs, wherein we examine the ability of the NODE approach to effectively learn the temporal dynamics and interpolate in the parameter space to make predictions for the unseen test case. We are particularly interested in exploring the ability of the continous-time, generative Latent ODE model \cite{rubanova2019latent} within the NODE approach, in which the ODE integration occurs in the latent space.
\vspace{-6mm}
\subsection{Contributions of this work}
 In this work,  we  model the temporal evolution of the turbulence prognostics such as kinetic energy, $k$, and its dissipation rate, $\epsilon$, which has emerged as a common reduced system to dynamically model turbulence ~\citep{jones72}.  
 Ground truth data is generated by extracting $k$ and $\epsilon$ from a series of DNS datasets, as described in section 2.
 We then apply the continuous-time NODE framework to model these time series ~\citep{chen2018neural}. 
 Compared to a discrete-depth network, such as a recurrent neural network (RNN), a NODE model can learn trajectories which may be sampled at arbitrary frequencies with standard ODE solvers ~\citep{chen2018neural,tan2018infocnf,rubanova2019latent} and hence is particularly well suited for physics problems. 
 
 Results from our experiments indicate that the NODE approach outperforms predictions from existing state-of-the-art models ~\cite{perot06} in the settings we consider. We observe NODE models to consistently remain within 1-2\% error with respect to the DNS solutions, almost two orders of magnitude less than state-of-the-art models. With these simple but fundamental experiments, we suggest that an automatic approach to training phenomenological reduced models for more complex turbulence phenomena, such as is necessary to integrate into existing climate models, is both practical and effective.

\section{Problem Background}
A common turbulence model solves for the turbulent kinetic energy $k$ and its dissipation rate $\epsilon$ \citep{jones72}, as:
\begin{align}
\label{eq:k1}
    \frac{D k}{D t}
    & = 
    \nabla \cdot \nu \nabla k+ \nabla \cdot \frac{\nu_t}{\sigma_k} \nabla k + \nu_t \left( \nabla \vec{u} + \nabla \vec{u}^{T} \right ) : \nabla \vec{u} - \epsilon \\
    \label{eq:k2}
    \frac{D \epsilon}{D t}
    & = 
    \nabla \cdot \nu \nabla \epsilon + \nabla \cdot \frac{\nu_t}{\sigma_\epsilon} \nabla \epsilon + \frac{\epsilon}{k}C_{e1} \nu_t \left( \nabla \vec{u} + \nabla \vec{u}^{T} \right ) : \nabla \vec{u} - \frac{\epsilon}{k}C_{e2} \epsilon
\end{align}
where $\nu$ is the fluid kinematic viscosity, $\sigma_k, \sigma_\epsilon, C_{e1}, C_{e2}$ are model constants, and $\nu_t = C_\mu k^2 / \epsilon$ is the turbulent viscosity 
where $C_\mu$ is another model constant.
Here, most constants can be derived from asymptotic analysis in low- or high-Reynolds number limits,
where the Reynolds number is a measure of turbulent scale separation, defined,
\begin{equation}
\mathrm{Re} \equiv \frac{k^2}{\nu \epsilon} \; .
\end{equation}
At intermediate Reynolds numbers, where turbulence transverses asymptotic limits, there is little theoretical justification for appropriate constants, yet it is well-accepted that $C_{e 2}$ has functional dependence on large-scale turbulence properties and $\mathrm{Re}$  when $1<\mathrm{Re}<100$ ~\citep[see][for a review]{perot06}. The case of the parameter $C_{e2}$ is particularly interesting, because under assumptions of homogeneity and isotropy, equations \eqref{eq:k1} and \eqref{eq:k2} are reduced 
to two coupled ODEs, wherein $C_{e2}$ is the only model parameter and the system is defined by:
\begin{equation}
    \frac{d k}{d t}
     = - \epsilon \; \text{ and }
    \frac{d \epsilon}{d t}
     =  - C_{e2}\frac{\epsilon^2}{k} \; .
    \label{eq:hi}
\end{equation}
Despite the simplicity of this reduced model, 
modeling the parameter $C_{e2}$
as a function of the large-scale fluid state is still an active research area for parameter regimes outside of asymptotic limits.  
A statistical description of turbulence at large length-scales is the power-law behaviour of kinetic energy $\hat k$ at small Fourier wavenumber,
\begin{equation}
\hat k (|\xi|) \propto |\xi|^{p_0} \; \text{ for } |\xi| \lessapprox \xi_l 
\end{equation}
where $\hat k$ is a function of  
the $L2$ norm of the wavenumber vector $\xi$, and $\xi_l$ is the integral wavenumber. Indeed,
the selection of $C_{e2}$ is known to significantly depend on the parameter $p_0$ \citep{saffman67,batchelor48}, leading to  
complex and nonlinear functional forms for $C_{e2}$ which adjust the model \eqref{eq:hi} to arbitrary values of $p_0$ (such as proposed by \citep{perot06}).  These models are enabled by constants and functional forms derived from data and regression from high-fidelity simulation. Given the dependence of these canonical models on high-fidelity turbulence data, it is reasonable to approach the prognostic modeling problem from a purely data-driven perspective with physics-inspired constraints for efficient and tractable numerical simulation.  


\section {Models}

\subsection{Direct Numerical Simulation}
In order to evaluate the tractability of modeling turbulent decay by data-driven methods, we develop a series of numerical experiment by simulating the governing equations of fluid motion, the Navier-Stokes equations, directly with robust Fourier spectral method ~\citep[c.f.][]{orszag72}.

\begin{figure}
  \begin{center}
    \includegraphics{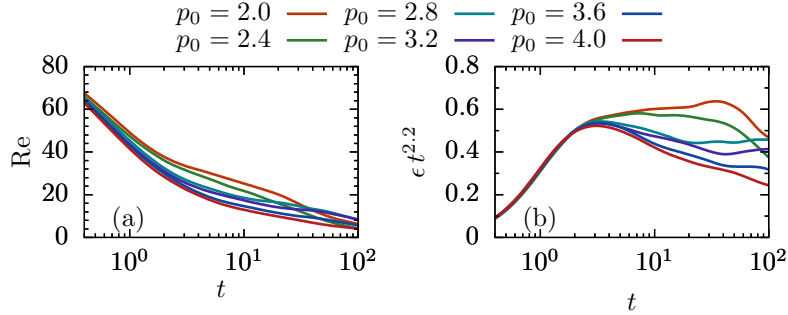}
  \end{center}
  \caption{Evolution of (a) the Reynolds number and (b) the kinetic energy dissipation rate, scaled by theoretical decay exponents corresponding to $p_0=2$, for single realizations of the cases described in Table \ref{tab:my_label}. Note the strong dependence of the evolution of $\epsilon$ on $\mathrm{Re}$ and $p_0$ after $t\approx2$.}
\label{fig:dns_decay}
\end{figure}

Turbulence is initialized by forcing the low-wavenumber power spectra at various $p_0$ to an appropriate form via a linear scheme suggested by~\citep{overholt98} and for the parameters summarized in Table \ref{tab:my_label}.  The parameter space has been designed to investigate Reynolds number regimes where existing models poorly calibrate and a range of $p_0$ observed in geophysical and engineering flows.

\begin{wraptable}[11]{R}{6.5cm}
\setlength{\columnsep}{0pt}%
\centering
\begin{tabular}{ccccc}
Case & $Re_0$ & $p_0$ & $N$\\
\hline   
Re100P20 &  99.85 &   2.0 &  5.5$\times 10^6$\\ 
Re100P24 &  99.61 &   2.4 &  7.1$\times 10^6$\\
Re100P28 &  100.7 &   2.8 &  9.0$\times 10^6$\\
Re100P32 &  100.2 &   3.2 &  1.1$\times 10^7$\\
Re100P36 &  100.3 &   3.6 &  1.4$\times 10^7$\\
Re100P40 &  100.4 &   4.0 &  1.7$\times 10^7$
\end{tabular}
\caption{Simulation database, each case is simulated with three realizations. The number of datapoints in each simulation is denoted by $N$, each simulation is run for over 5,000 timesteps.}
\label{tab:my_label}
\end{wraptable}
Once the flows are statistically stationary, the forcing is turned off and the turbulence is allowed to decay. The process is repeated for three realization of each parameter combination in order to sample the data variability.

Trajectories of the Reynolds number and $\epsilon$ are shown in figure \ref{fig:dns_decay}, where time has been appropriately non-dimensionalized by a turbulence time-scale. Our objective is to discover a $\mathrm{Re}$-dependent model with $p_0$ as a parameter which outperforms existing state-of-the-art models.

\subsection {Neural ODE for time-series}

We use the continuous-time, generative Neural ODE approach called Latent ODE ~\citep[][section 5]{chen2018neural}, to model the turbulent kinetic energy dissipation from the DNS data presented in the previous section. In the context of time-series, the model represents each observation by a latent trajectory, $z_t$, determined by


\begin{equation}
\frac{d z_t}{dt} = f(z_t,\phi)
\end{equation}



where $f$ specifies the dynamics of the hidden state of the network itself and $\phi$ is the set of neural network parameters. The model has three different parts to it. First, a recognition recurrent neural network (RNN) reads the observations from the DNS data backwards in time in order to determine an initial latent representation, $z_{t0}$ for each observation ($k$ and $\epsilon$). Second, this initial latent representation is used as an input to the ODE solver together with a function $f$, parameterized by a neural network, and is used to obtain latent space observations at all given times: 



\begin{equation}
\label{eq:odesolve}
z_{t_1}, z_{t_2}, z_{t_3}, ....,z_{t_N} = \mathrm{ODESolve} ~(z_{t_0}, f, \theta_f, t_{0}, t_{1}, t_{2},...., t_{N})
\end{equation}

In the third and final step, the latent space observations are decoded back to data space by another network. Training is performed by optimizing the parameters of the networks at each part (recognition RNN, ODE function, and decoder) in order to minimize the error between input and output or, in the context of a variational autoencoder~\citep{kingma2013auto,rezende2014stochastic}, maximize the evidence lower bound (ELBO).



One of the main advantages of the latent ODE approach in the prediction of time series data lies in the fact that the previously shown function $f$ is time invariant. This means that given a latent state, $z_t$, one can define a unique latent trajectory. Therefore, a model trained to fit any given set of observations is able to extrapolate the latent trajectory arbitrarily far forwards (forecasting), or backwards in time~\citep{rubanova2019latent}. Alternatively, one could also train the model on a given set of input parameters and generalize a solution to a new parameter value not seen during training.

In this work, we tested a latent ODE model architecture for deriving a turbulent kinetic energy dissipation model sensible to various $p_0$ values. The RNN encoder has 25 hidden units and 4 units in the latent space. The function $f$ is parameterized with a fully-connected (FC) layers network with one hidden layer containing 20 hidden units. The decoder has a similar network, also with 20 hidden units. We train the model with DNS results on a given set of large scale turbulence properties (such as $p_0$) and attempt to generalize a solution for a new set of unseen test conditions.

\section{Experimental Results}
\begin{figure}
  \begin{center}
    \includegraphics{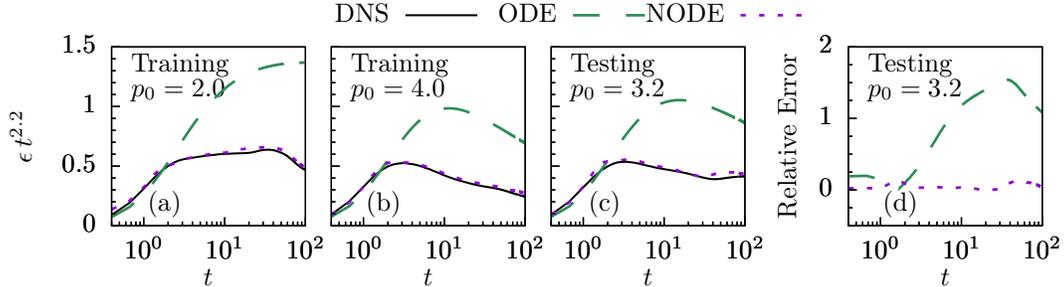}
  \end{center}
  \vspace{-3mm}
  \caption{Comparison of models for the evolution of dissipation rate, scaled as in figure \ref{fig:dns_decay}. Panels (a) and (b) show the purely data-driven model, which accounts for changes in $\mathrm{Re}$ and $p_0$, to capture the trends of the ground truth much closer than the state-of-the-art analytic model ~\citep{perot06}. The same is observed for the testing case in panel (c), whereas panel (d), which quantifies the error with respect to the ground truth data, illustrates that the NODE predictions stay within 1-2\% of the DNS solutions.}
\label{fig:results}

\end{figure}

Data-driven models are compared to state-of-the-art analytic models ~\citep{perot06} in figure \ref{fig:results}.  Recalling that our objective is to capture transient $\mathrm{Re}$ and $p_0$ dependence in the evolution of turbulence prognostics, we compare against a model which accounts for these two parameters with complex and non-linear sub-models.  We have selected an intermediate case Re100P32 for testing for the results presented here, but observed qualitatively similar behaviour when other cases have been held-out in cross-validation. 
The training sets are shown in figures \ref{fig:results}a,b, which illustrate the capability of the Neural ODE model applied to time-series data to represent the evolution of the turbulent kinetic energy dissipation rate.  The performance of the purely data-driven approach for the testing case is shown in figures \ref{fig:results}c,d.  We observe that while both models accurately predict the initial period of decay for $t\lessapprox 2$, the analytic model over-estimates the dissipation rate for later times.   We believe these observations are due to the analytic ODE model poorly capturing low Reynolds number effects in the final stages of decay (see figure \ref{fig:dns_decay}a). In comparison, the purely data-driven Neural ODE approach accounts for this transient such that it more accurately forecasts the dissipation rate at later times.

\section{Conclusions and future applications}
We have proposed a well-constrained, fundamental physics problem for the application of data-driven modeling via Neural ODEs.  By comparison with state-of-the-art models for the equivalent problem, we have demonstrated that the Neural ODE model more accurately predicts the evolution of the dissipation rate at multiple forcing objectives (as defined by $p_0$), which is a testament to the generalizability of the framework. This result is encouraging for the development of models for more complex turbulence phenomena, such as model flow configuration relevant to geophysical turbulence as used in climate models, where the evolution of additional prognostics are necessary to model.  These prognostics, such as the flux energy between kinetic energy and gravitational potential energy reservoirs, are currently poorly parameterized ~\cite{ivey18} and outperforming data-driven approaches have an opportunity to immediately benefit such existing models. Another area of future investigation is to explore the interpretability \cite{dong2019geometrization, ribeiro2016model, guidotti2019survey} of these \textit{black-box} ML models, such they can be better generalized to more complex physics problems.





\section{Acknowledgements}
The research activities of the authors the from Los Alamos National Laboratory are supported by a LDRD project "Machine Learning for Turbulence (MELT)" \#20190059DR. 

The data utilized in this study was generated for related research sponsored by the Department of Energy NNSA Advanced Simulation and Computing (ASC) program through the Physics and Engineering Models -- Mix \& Burn (PEM-M\&B) and the Advanced Technology Development and Mitigation -- Machine Learning (ATDM-ML) projects.

The authors from the University of Massachusetts would like to acknowledge the ICEnet consortium for the research funding and the GPU resources. Support from the members of the consortium, including NVIDIA Corp., SIEMENS, The MathWorks Inc., Cummins, Convergent Science Inc., \& AVL, is appreciated.


This document has been approved by Los Alamos National Laboratory for public release as LA-UR-19-29250



\begin{thebibliography}{10}

\bibitem{portwood2019physics}
Gavin Portwood, Misha Chertkov, Balasubramanya Nadiga, Juan Saenz, and Daniel
  Livescu.
\newblock Physics-informed deep neural networks applied to scalar subgrid flux
  modeling in a mixed {DNS}/{LES} framework.
\newblock {\em Bulletin of the American Physical Society}, 2019.

\bibitem{portwood2019autonomous}
Gavin Portwood, Juan Saenz, and Daniel Livescu.
\newblock Autonomous {RANS}/{LES} hybrid models with data-driven subclosures.
\newblock {\em Bulletin of the American Physical Society}, 2019.

\bibitem{mitra2019data}
Peetak Mitra, Mateus Dias~Ribeiro, and David Schmidt.
\newblock A data-driven approach to modeling turbulent flows in an engine
  environment.
\newblock {\em Bulletin of the American Physical Society}, 2019.

\bibitem{chen2018neural}
Tian~Qi Chen, Yulia Rubanova, Jesse Bettencourt, and David~K Duvenaud.
\newblock Neural ordinary differential equations.
\newblock In {\em Advances in neural information processing systems}, pages
  6571--6583, 2018.

\bibitem{tan2018infocnf}
Tan~M. Nguyen, Animesh Garg, Richard~B. Baraniuk, and Anima Anandkumar.
\newblock {InfoCNF}: An efficient conditional continuous normalizing flow with
  adaptive solvers.
\newblock {\em ICML 2019 Workshop on Invertible Neural Networks and Normalizing
  Flows}, 2019.

\bibitem{rubanova2019latent}
Yulia Rubanova, Ricky~TQ Chen, and David Duvenaud.
\newblock Latent {ODEs} for irregularly-sampled time series.
\newblock {\em arXiv preprint arXiv:1907.03907}, 2019.

\bibitem{sun2019neupde}
Yifan Sun, Linan Zhang, and Hayden Schaeffer.
\newblock Neupde: Neural network based ordinary and partial differential
  equations for modeling time-dependent data.
\newblock {\em arXiv preprint arXiv:1908.03190}, 2019.

\bibitem{maulik2019time}
Romit Maulik, Arvind Mohan, Bethany Lusch, Sandeep Madireddy, and Prasanna
  Balaprakash.
\newblock Time-series learning of latent-space dynamics for reduced-order model
  closure.
\newblock {\em arXiv preprint arXiv:1906.07815}, 2019.

\bibitem{san2019artificial}
Omer San, Romit Maulik, and Mansoor Ahmed.
\newblock An artificial neural network framework for reduced order modeling of
  transient flows.
\newblock {\em Communications in Nonlinear Science and Numerical Simulation},
  77:271--287, 2019.

\bibitem{dias2019data}
Mateus Dias~Ribeiro, Gavin~D Portwood, Peetak Mitra, Tan Mihn~Nyugen,
  Balasubramanya~T Nadiga, Michael Chertkov, Anima Anandkumar, and David~P
  Schmidt.
\newblock A data-driven approach to modeling turbulent decay at non-asymptotic
  reynolds numbers.
\newblock {\em Bulletin of the American Physical Society}, 2019.

\bibitem{jones72}
WP~Jones and Brian~Edward Launder.
\newblock The prediction of laminarization with a two-equation model of
  turbulence.
\newblock {\em International journal of heat and mass transfer},
  15(2):301--314, 1972.

\bibitem{perot06}
J.~B. Perot and S.~M. de~Bruyn~Kops.
\newblock Modeling turbulent dissipation at low and moderate {R}eynolds
  numbers.
\newblock {\em J. Turbulence}, 7:1--14, 2006.

\bibitem{saffman67}
P.G. Saffman.
\newblock The large-scale structure of homogeneous turbulence.
\newblock {\em J. Fluid Mech.}, 27(3):581--593, 1967.

\bibitem{batchelor48}
G.~K. Batchelor and A.~A. Townsend.
\newblock Decay of turbulence in the final period.
\newblock {\em P. Roy. Soc. Lond. A. Mat.}, 194:527, 1948.

\bibitem{orszag72}
S.~A. Orszag and G.~S. Patterson.
\newblock Numerical simulation of turbulence.
\newblock In M.~Rosenblatt and C.~{Van Atta}, editors, {\em Statistical Models
  and Turbulence}, volume~12 of {\em Lecture Notes in Physics}, pages 127--147.
  Springer, New York, 1972.

\bibitem{overholt98}
M.~R. Overholt and S.~B. Pope.
\newblock A deterministic forcing scheme for direct numerical simulations of
  turbulence.
\newblock {\em Comput. Fluids}, 27:11--28, 1998.

\bibitem{kingma2013auto}
Diederik~P Kingma and Max Welling.
\newblock Auto-encoding variational bayes.
\newblock {\em arXiv preprint arXiv:1312.6114}, 2013.

\bibitem{rezende2014stochastic}
Danilo~Jimenez Rezende, Shakir Mohamed, and Daan Wierstra.
\newblock Stochastic backpropagation and approximate inference in deep
  generative models.
\newblock {\em arXiv preprint arXiv:1401.4082}, 2014.

\bibitem{ivey18}
Gregory~N Ivey, Cynthia~E Bluteau, and Nicole~L Jones.
\newblock Quantifying diapycnal mixing in an energetic ocean.
\newblock {\em J. Geophys. Res.-Oceans}, 123(1):346--357, 2018.

\bibitem{dong2019geometrization}
Xiao Dong and Ling Zhou.
\newblock Geometrization of deep networks for the interpretability of deep
  learning systems.
\newblock {\em arXiv preprint arXiv:1901.02354}, 2019.

\bibitem{ribeiro2016model}
Marco~Tulio Ribeiro, Sameer Singh, and Carlos Guestrin.
\newblock Model-agnostic interpretability of machine learning.
\newblock {\em arXiv preprint arXiv:1606.05386}, 2016.

\bibitem{guidotti2019survey}
Riccardo Guidotti, Anna Monreale, Salvatore Ruggieri, Franco Turini, Fosca
  Giannotti, and Dino Pedreschi.
\newblock A survey of methods for explaining black box models.
\newblock {\em ACM computing surveys (CSUR)}, 51(5):93, 2019.

\end{thebibliography}

\end{document}